\documentclass[conference,a4paper]{IEEEtran}
\IEEEoverridecommandlockouts
\usepackage{cite}
\usepackage{amsmath,amssymb,amsfonts}
\usepackage{algorithmic}
\usepackage{graphicx}
\usepackage{textcomp}
\usepackage{xcolor}
\usepackage{subfigure}

\def\BibTeX{{\rm B\kern-.05em{\sc i\kern-.025em b}\kern-.08em
    T\kern-.1667em\lower.7ex\hbox{E}\kern-.125emX}}
\begin{document}

\title{Generalizable Variable-Rate Adaptation \\ for Image Coding for Machines}

\author{\IEEEauthorblockN{Yui Tatsumi}
\IEEEauthorblockA{\textit{Graduate School of FSE,} \\
\textit{Waseda University}\\
Tokyo, Japan \\
yui.t@fuji.waseda.jp}
\and
\IEEEauthorblockN{Hiroshi Watanabe}
\IEEEauthorblockA{\textit{Graduate School of FSE,} \\
\textit{Waseda University}\\
Tokyo, Japan \\
hiroshi.watanabe@waseda.jp}
}

\maketitle

\begin{abstract}
Image Coding for Machines (ICM) aims to preserve task-relevant information for machine vision under limited bitrates. However, most ICM models are trained for a fixed operating point, and extending them to variable-rate compression is challenging because their training objectives differ across tasks and model designs. In this paper, we propose a loss-agnostic distillation framework for variable-rate ICM. A pretrained fixed-rate ICM model is used as a frozen teacher, while a student model learns variable-rate behavior through importance-adaptive quantization and distillation-based optimization. This enables continuous bitrate control with a single model while preserving task-relevant latent components, without requiring the original task-specific loss or downstream task labels. Experiments based on existing ICM models demonstrate that the proposed method provides continuous bitrate control and achieves competitive recognition performance compared with the fixed-rate models.
\end{abstract}

\begin{IEEEkeywords}
Image coding for machines, variable-rate
\end{IEEEkeywords}

\section{Introduction}
Learned Image Compression (LIC) has attracted increasing attention due to its superior rate-distortion performance over traditional codecs such as JPEG\cite{JPEG} and VVC-Intra\cite{VVC}. In recent years, images are increasingly analyzed by machine vision models in applications such as traffic monitoring and smart agriculture. In these scenarios, perceptual quality is not always essential; instead, compressed representations should preserve information relevant to downstream recognition tasks. This has motivated the development of Image Coding for Machines (ICM).

In practical ICM scenarios, the required bitrate may vary depending on network conditions and application needs. However, many learned compression models operate only at fixed operating points and require separate trained models for different bitrate targets. This increases both storage and training costs. To address this issue, various variable-rate LIC methods have been proposed for human-oriented compression. In contrast, variable-rate control for ICM remains underexplored. Since human- and machine-oriented coding have different optimization objectives, ICM requires a dedicated variable-rate approach.

A key challenge in developing variable-rate ICM is that existing ICM models are trained with diverse objectives, such as task loss, semantic supervision, and task-relevant latent preservation. While this diversity is a strength of ICM, it hinders the application of variable-rate mechanisms that depend on a specific training objective across different ICM models.

In this paper, we propose an importance-adaptive distillation-based variable-rate framework for ICM. The proposed method uses a pretrained fixed-rate ICM model as a teacher and trains a student model to preserve its learned compression characteristics while enabling bitrate control. We use the estimated entropy-model scale parameters as latent importance and adapt both the quantization step size and the distillation loss accordingly. This allows the student to preserve task-relevant latent components more effectively at low bitrates. Because the framework relies only on the outputs of a pretrained ICM model, it can be applied to different ICM architectures without additional task labels, saliency maps, or modifications to downstream recognition models.

\begin{figure*}[t]
\centering
\includegraphics[width=0.96\linewidth]{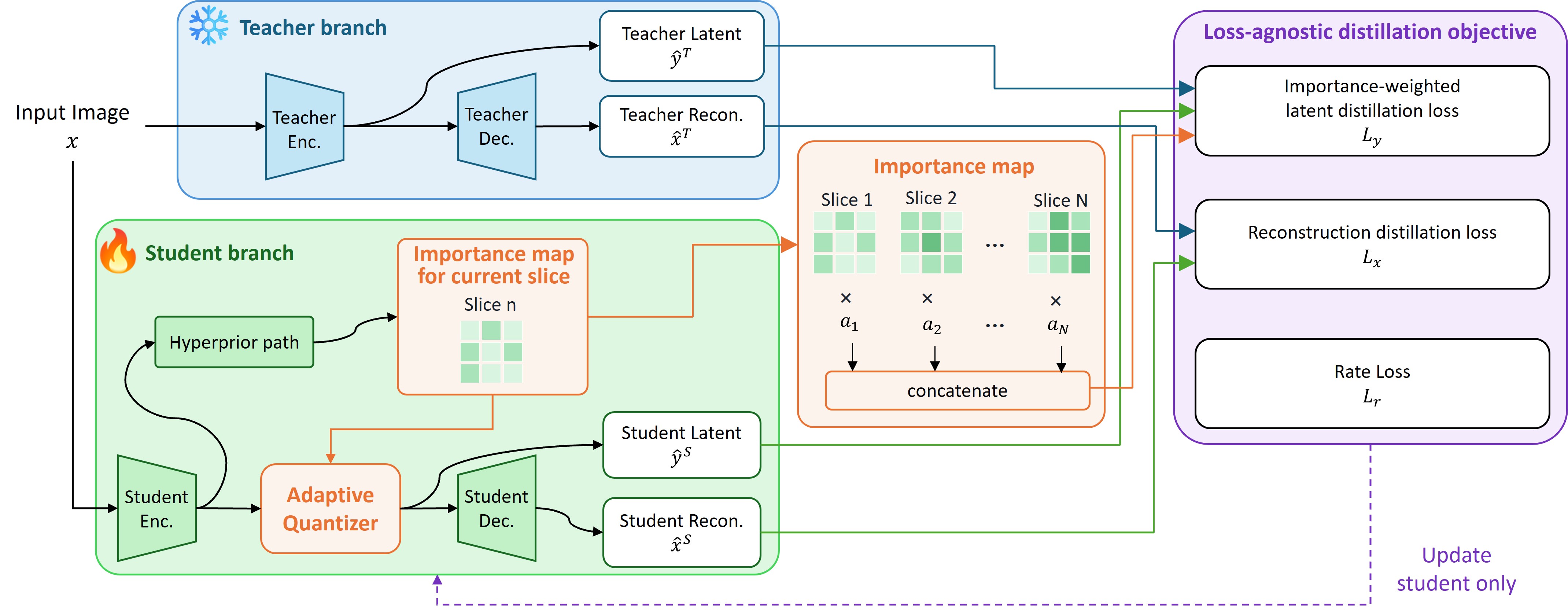}
\caption{Overview of the proposed training framework.}
\label{fig:overview}
\end{figure*}

\section{Related Work}
\subsection{Image Coding for Machines}
Early LIC methods mainly target human-oriented compression\cite{balle}-\cite{MLIC}. Most adopt the hyperprior architecture introduced by J. Ballé \textit{et al.}\cite{balle} to model the distribution of quantized latents.
Autoregressive entropy models further improve the performance by exploiting
various dependencies such as spatial\cite{minnen} and channel-wise\cite{ch-arm}.
These methods are typically trained with a rate-distortion loss, where $\lambda$ controls the trade-off between bitrate and reconstruction distortion. Since $\lambda$ is fixed during training, separate models must be trained and stored to support different bitrate targets.

Unlike human-oriented compression, ICM adopts diverse optimization objectives for machine vision tasks. Task loss-based approaches directly optimize downstream performance, such as object detection or segmentation\cite{task-first}-\cite{task-last}, while ROI-based methods allocate more bits to task-relevant regions using analysis-driven importance maps\cite{ROI-first}-\cite{ROI-last}. More recently, task-generalized ICM methods have been developed to reduce dependence on specific downstream tasks\cite{SA-ICM}-\cite{Delta-ICM}. For instance, SA-ICM\cite{SA-ICM} preserves object boundary information for task-agnostic machine analysis, whereas UG-ICM\cite{UG-ICM} uses CLIP\cite{CLIP}-based supervision and preference-conditional decoding to support both human perception and generalized machine analytics with a unified bitstream.

\subsection{Variable-rate Compression}
Variable-rate LIC has been widely studied for human-oriented compression\cite{vbr-RNN}-\cite{Selective}, but remains underexplored for ICM. M. Song \textit{et al.}~\cite{Qmap} control spatial bit allocation using a quality map. Although adaptable to task-aware compression, the codec itself is trained for image reconstruction rather than machine-oriented objectives. J. Liu \textit{et al.}~\cite{ECCV2024} propose a rate-distortion-cognition controllable codec, but it relies on a dedicated architecture and cognition-oriented objective. Thus, these methods do not provide a general solution for diverse ICM models with different optimization objectives. AQVR-ICM~\cite{AQVR-ICM} enables training-free variable-rate control of pretrained ICM models using entropy-model scale parameters without retraining. However, since the model is not optimized for variable-rate operation, performance is limited, particularly at low bitrates. These limitations motivate a trainable, loss-agnostic variable-rate framework applicable to pretrained ICM models with diverse objectives.

\subsection{Knowledge Distillation}
Knowledge distillation has been widely utilized to transfer model capabilities from larger, high-performing teacher models to compact student models, mainly for model compression and efficient inference\cite{Hinton}. 
Recent studies have also applied distillation to LIC for reducing the computational complexity while preserving rate-distortion performance\cite{distillation-LIC}.
However, most existing distillation-based compression methods focus on model acceleration or rate-distortion preservation, while little attention has been paid to using distillation for other objectives such as generalized variable-rate adaptation in ICM.

\section{Proposed Method}
\subsection{Overall Framework}
In this paper, we propose an importance-adaptive and loss-agnostic framework for variable-rate ICM. As illustrated in Fig. \ref{fig:overview}, the framework adapts a pretrained fixed-rate ICM model to support variable-rate compression while preserving the behavior learned from its original optimization objective.

The proposed framework consists of two components. 
First, an adaptive quantizer adjusts the quantization step size based on an importance map derived from entropy-model scale parameters, enabling bitrate control while preserving task-relevant latent components. 
Second, a loss-agnostic distillation objective enables adaptation to different pretrained ICM models without requiring task-specific losses or recognition labels. The latent distillation loss is further weighted by the importance map to prioritize task-relevant latent components and mitigate recognition performance degradation.
\begin{figure*}[t]
\centering
\includegraphics[width=\linewidth]{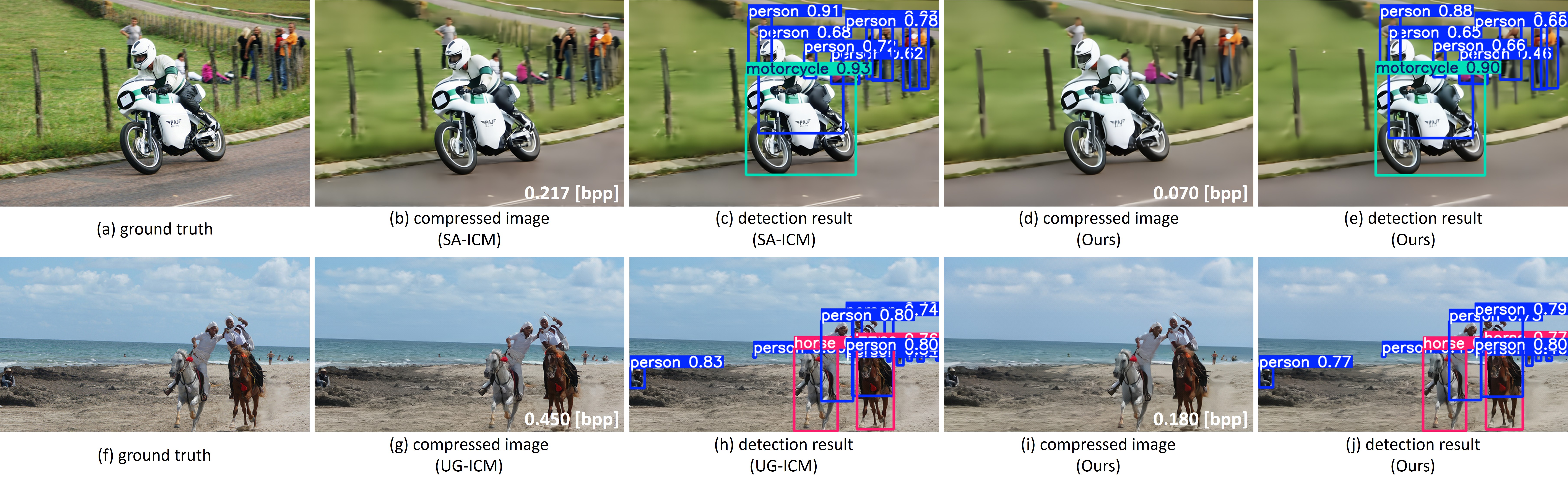}
\caption{Examples of decoded images and YOLOv5 object detection results.}
\label{fig:outputs}
\end{figure*}

\subsection{Scale-Aware Importance Estimation}
\label{section:importance-map}
The proposed method estimates latent importance from entropy-model scale parameters. In hyperprior-only models, scale parameters are obtained directly from the decoded hyper-latent, whereas autoregressive models predict them sequentially according to the decoding order. Accordingly, the importance estimation must follow the corresponding autoregressive structure. Here, we describe the formulation for a channel-wise autoregressive model, which can be extended to spatial autoregressive models by following their decoding order.

In Sections~\ref{section:importance-map} and~\ref{section:quantization}, all latent variables and entropy-model parameters are obtained from the student model. In a channel-wise autoregressive entropy model, the latent representation is split into $N$ slices along the channel dimension and processed sequentially. For the $n$-th slice $(n = 1, 2, \ldots, N)$, the entropy model predicts the mean and scale parameters, and we use the scale parameter $\sigma^{(n)}$ to estimate the importance of each latent component.
Since both the encoder and decoder predict $\sigma^{(n)}$ from the decoded hyper-latent \(\hat{z}\) and previously reconstructed slices following the same autoregressive order, the same importance map and quantization step size can be obtained without transmitting an additional importance map.
We first compute a stabilized scale representation:
\begin{equation} 
s^{(n)}= \log \left( 1 + \left| \sigma^{(n)} \right| \right). 
\end{equation} 
To capture both channel-wise and spatial tendencies, we compute the spatially averaged  channel-wise scale score $\bar{s}^{(n)}_{c}$, the channel-averaged spatial scale score $\bar{s}^{(n)}_{h,w}$, and the global average $\bar{s}^{(n)}$ of $s^{(n)}$. The slice-wise importance map is then defined as 
\begin{equation} 
I^{(n)}_{c,h,w} = \frac{\bar{s}^{(n)}_{c}}{\bar{s}^{(n)}+\epsilon} \cdot \frac{\bar{s}^{(n)}_{h,w}}{\bar{s}^{(n)}+\epsilon}. 
\label{eqn:slice-wise_importance}
\end{equation} 

The slice-wise importance map $I^{(n)}$ defined in Eq. (\ref{eqn:slice-wise_importance}) is used for adaptive quantization of the current slice. In the autoregressive entropy model, the relative importance among slices cannot be determined at that stage.
For latent distillation, however, all slice-wise scale scores are available after processing the full latent representation. 
Therefore, we further introduce a slice-level weight for the distillation loss:
\begin{equation}
a_n
=
\frac{
\bar{s}^{(n)}
}{
\frac{1}{N}
\sum_{n'=1}^{N}
\bar{s}^{(n')}
+
\epsilon
}.
\end{equation}
Then, the importance map for the loss function is obtained by concatenating the weighted slice-wise maps:
\begin{equation} 
\label{eqn:distillation-imporance}
I = conc\left(a_1 I^{(1)}, a_2 I^{(2)}, \dots , a_N I^{(N)}\right).
\end{equation} 

\begin{figure*}
\centering
\subfigure[Object Detection by YOLOv5]{%
\includegraphics[width=0.32\hsize]{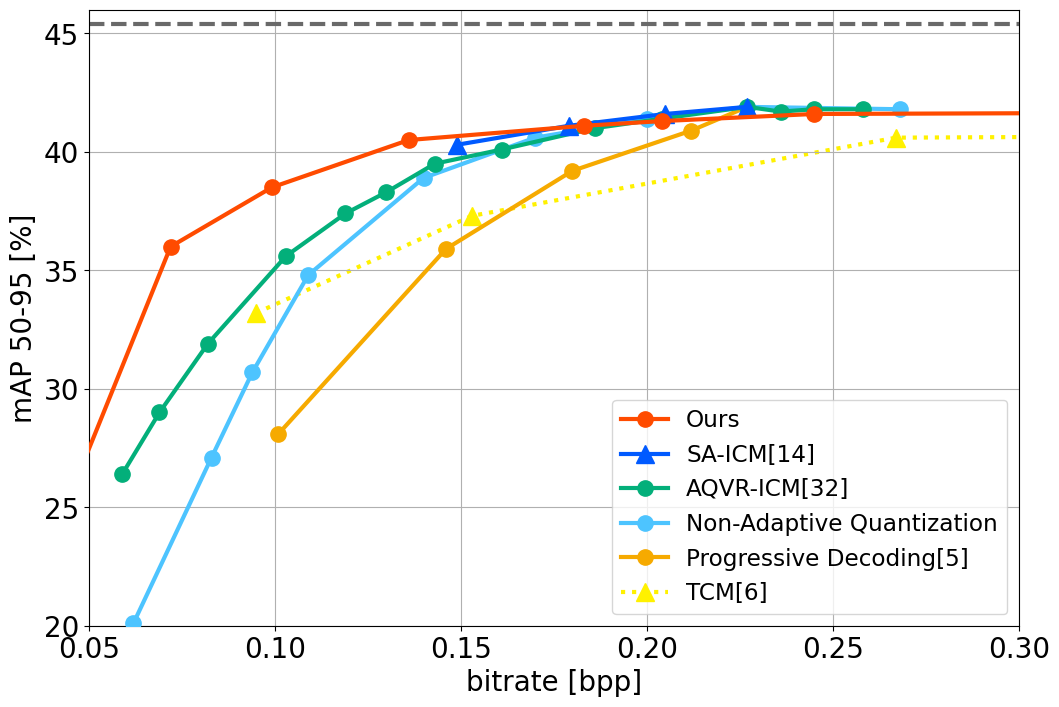}}%
\hspace{0.0025pt}
\subfigure[Object Detection by Mask R-CNN]{%
\includegraphics[width=0.32\hsize]{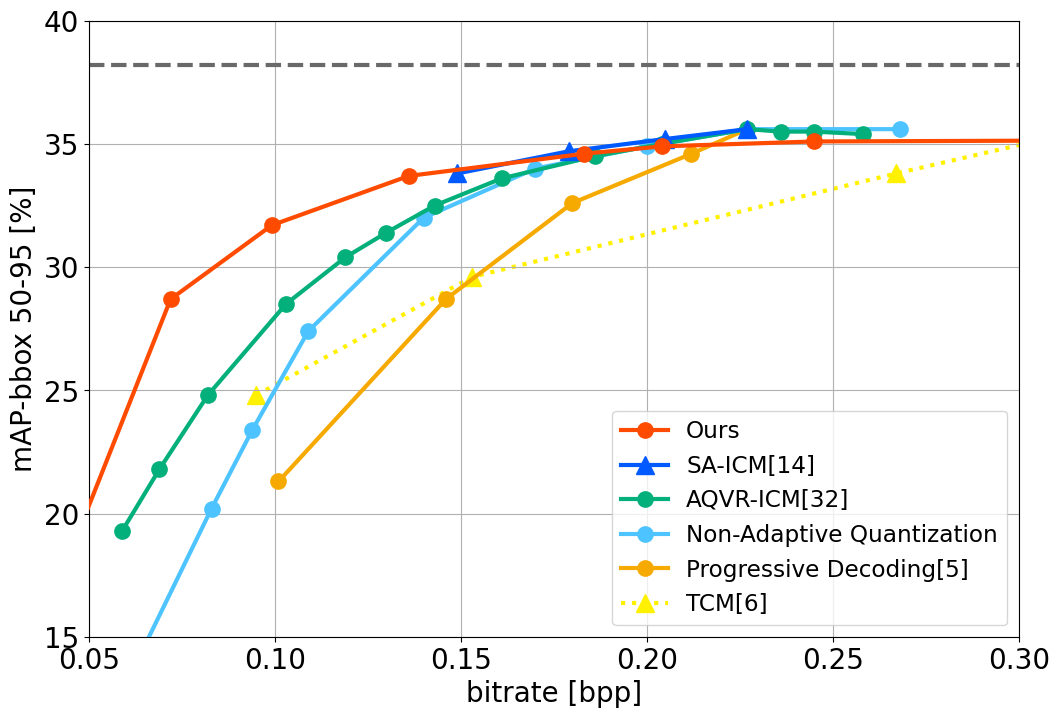}}%
\hspace{0.0025pt}
\subfigure[Instance Segmentation by Mask R-CNN]{%
\includegraphics[width=0.32\hsize]{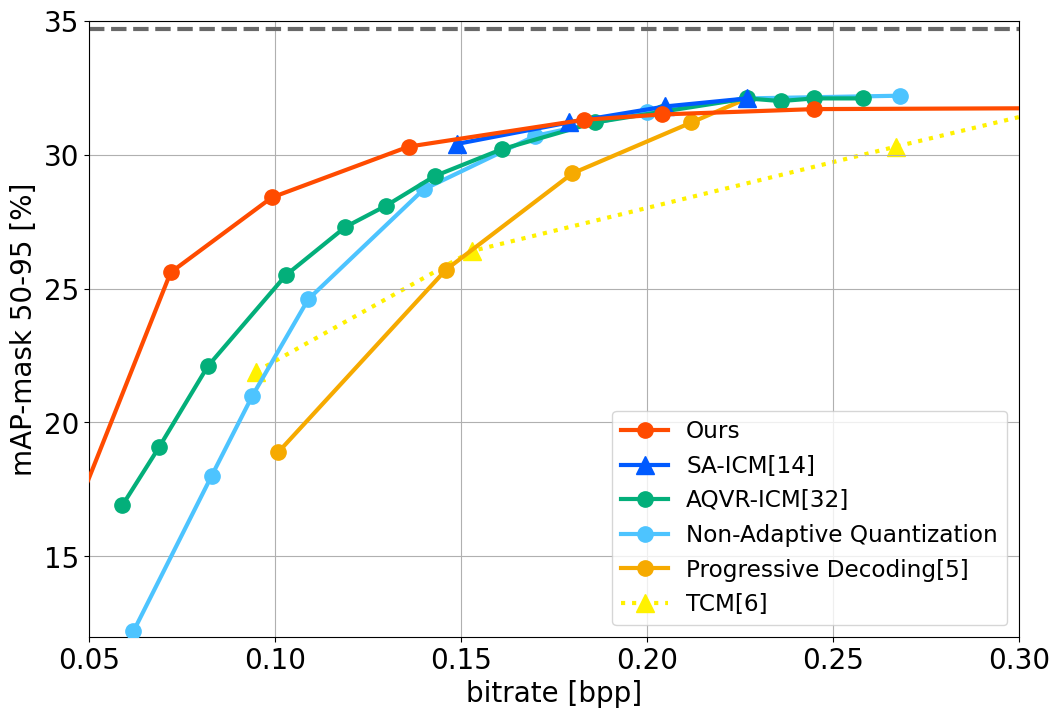}}%
\hspace{0.0025pt}
\caption{Image compression performance based on SA-ICM\cite{SA-ICM} for different recognition tasks. Solid/dotted curves denote machine-/human-oriented codecs, and circular/triangular markers denote variable-rate/fixed-rate results. The gray dotted lines represent the
results on uncompressed images.}
\label{fig:SA-ICM}

\end{figure*}
\begin{figure*}
\centering
\subfigure[Object Detection by YOLOv5]{%
\includegraphics[width=0.32\hsize]{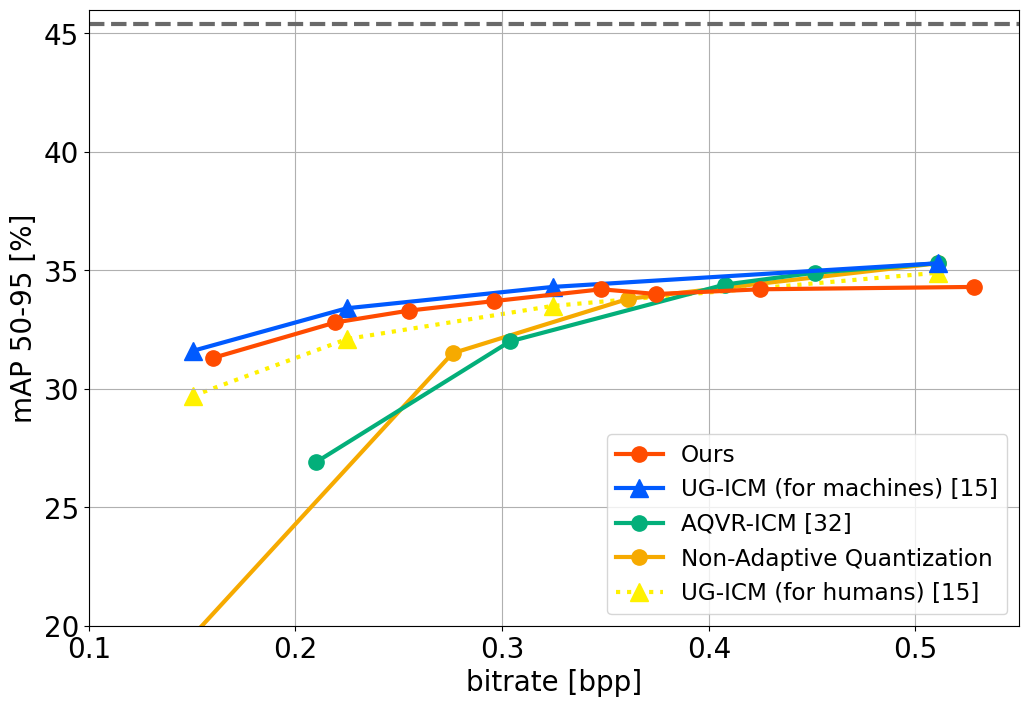}}%
\hspace{0.0025pt}
\subfigure[Object Detection by Mask R-CNN]{%
\includegraphics[width=0.32\hsize]{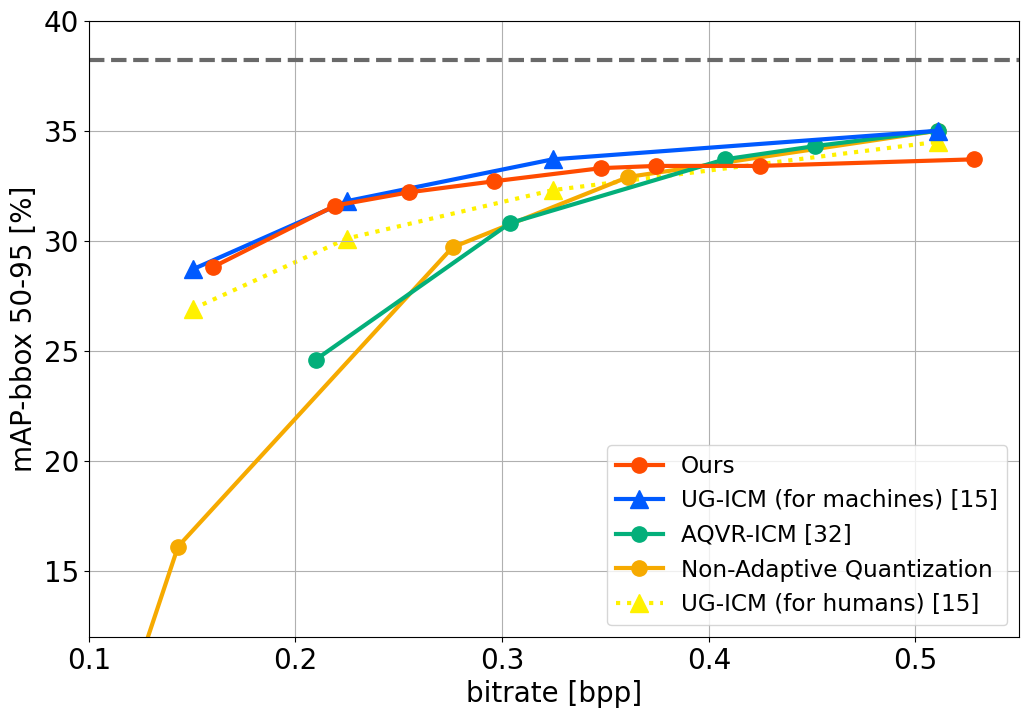}}%
\hspace{0.0025pt}
\subfigure[Instance Segmentation by Mask R-CNN]{%
\includegraphics[width=0.32\hsize]{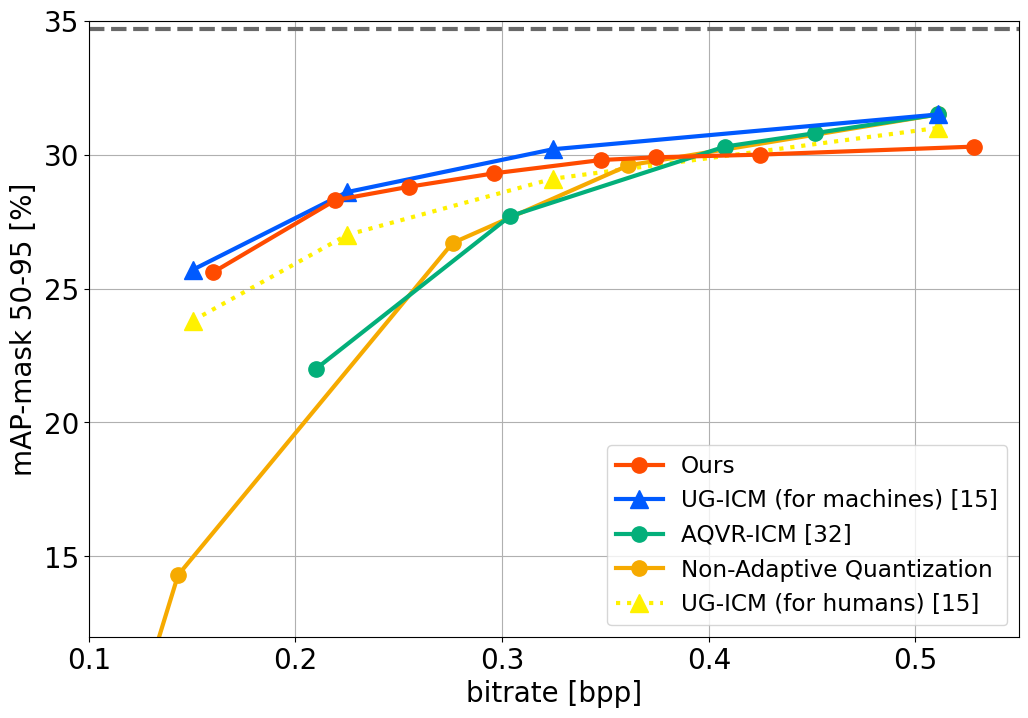}}%
\hspace{0.0025pt}
\caption{Image compression performance based on UG-ICM\cite{UG-ICM} for different recognition tasks. Solid/dotted curves denote machine-/human-oriented codecs, and circular/triangular markers denote variable-rate/fixed-rate results. The gray dotted lines represent the
results on uncompressed images.}
\label{fig:UG-ICM}
\end{figure*}

\subsection{Importance-Adaptive Variable-Rate Quantization}
\label{section:quantization}
We assign an adaptive quantization step size for each latent element in the current slice using the importance map $I^{(n)}$.
Let $\tilde{I}^{(n)} = I^{(n)}/(\bar{I}^{(n)}+\epsilon)$
denote the normalized slice-wise importance map. We first compute an importance-dependent relative quantization factor as
\begin{equation}
R^{(n)} = \left(\tilde{I}^{(n)}+\epsilon\right)^{-\gamma},
\end{equation}
where $\gamma$ controls the strength of importance-based bit allocation. 
A larger importance value gives a smaller relative factor, and therefore important latent components are quantized more finely.

To control the overall bitrate, we introduce a scalar rate-control parameter $\alpha$.
The global rate-control factor and the protection factor are defined as
\begin{equation}
G(\alpha)=\exp(k\alpha), \qquad
P(\alpha)=1-\exp(-c|\alpha|),
\end{equation}
where $k$ controls the bitrate sensitivity to $\alpha$, and $c$ controls how rapidly the protection factor $P(\alpha)$ increases as $\alpha$ moves away from zero. $P(\alpha)$ suppresses importance-adaptive quantization near the original teacher operating point.
The final adaptive quantization step size is given by
\begin{equation}
\Delta^{(n)} = G(\alpha)\left[1 + P(\alpha)\left(R^{(n)} - 1\right)\right].
\end{equation}
Varying $\alpha$ changes the global bitrate
while redistributing bits according to the importance map.

Given the latent representation $y^{(n)}$, the predicted mean $\mu^{(n)}$, and the adaptive step size $\Delta^{(n)}$ for the $n$-th slice, quantization is performed as
\begin{equation}
\hat{y}^{(n)}
=
\mathrm{round}
\left(
\frac{y^{(n)}-\mu^{(n)}}{\Delta^{(n)}}
\right)
\Delta^{(n)}
+
\mu^{(n)}.
\end{equation}
Since the quantization interval depends on $\Delta$, the likelihood is computed over an interval of width $\Delta_i$ for each latent element:
\begin{equation}
p_{\hat{y}|\hat{z}}(\hat{y}_{i}|\hat{z},\Delta_i)
=
\int_{\hat{y}_{i}-\Delta_i/2}^{\hat{y}_{i}+\Delta_i/2}
\mathcal{N}(t|\mu_{i},(\sigma_{i})^2)dt.
\end{equation}

We do not apply adaptive quantization to the hyperprior path because coarse 
quantization of the hyper-latent $z$ can degrade scale prediction, while $\hat{z}$ has little impact on overall rate control, as mentioned in previous studies\cite{F. Kamisli},\cite{AQVR-ICM}.

\subsection{Loss-Agnostic Distillation Objective}
The proposed method uses a distillation-based objective to train the student model without relying on the original ICM loss function.
Let $T$ and $S$ denote the teacher and student models, respectively. 
The training loss function combines latent distillation, reconstruction distillation, and rate loss:
\begin{equation}
\mathcal{L}
=
\lambda_y \mathcal{L}_y
+
\lambda_x \mathcal{L}_x
+
\lambda_r \mathcal{L}_r.
\label{eqn:loss}
\end{equation}

Using the normalized importance map
$\tilde{I}=I/(\bar{I}+\epsilon)$, the latent and reconstruction
distillation losses are given by
\begin{align}
    \mathcal{L}_y
    &=
    \mathbb{E}
    \left[
    \tilde{I}
    \left(
    \hat{y}^{S}
    -
    \hat{y}^{T}
    \right)^2
    \right], \\
    \mathcal{L}_x
    &=
    \mathbb{E}
    \left[
    \left(
    \hat{x}^{S}
    -
    \hat{x}^{T}
    \right)^2
    \right].
\end{align}
Here, $\mathcal{L}_r$ denotes the estimated bpp of the student
model computed from the entropy-model likelihoods of
$\hat{y}^{S}$ and $\hat{z}^{S}$.

\section{Experiment}
We use COCO-train dataset for training and COCO-val for evaluation\cite{COCO}. The student model is trained for 10 epochs with Adam optimizer, a learning rate of $1\times10^{-4}$, and a batch size of 8. For each base model, the highest-bitrate pretrained checkpoint is used as the fixed teacher, and the student is initialized from the same weights. The loss weights in Eq.~(\ref{eqn:loss}) are $\lambda_y=1.0$, $\lambda_x=0.03$, and $\lambda_r=0.05$. The rate-control parameter is sampled from $\alpha\in\{0,0.5,1.0,2.0\}$ during training and evaluated at $\alpha\in\{-0.5,0,0.3,0.5,1.0,1.5,2.0,3.0\}$. Reported bpp values are estimated from the entropy-model likelihoods of \(\hat{y}\) and \(\hat{z}\), excluding arithmetic-coding overhead and \(\alpha\) signaling. Since \(\alpha\) is a single scalar sent once per image, its signaling overhead is small.


The proposed method is applicable to any fixed-rate ICM model with a hyperprior structure. We validate it on two models with different
training objectives: SA-ICM~\cite{SA-ICM} and MLIC++~\cite{MLIC}-based UG-ICM~\cite{UG-ICM}. 
We use \((k,c,\gamma)=(1.0,1.0,0.5)\) for SA-ICM and \((0.35,0.25,0.25)\) for UG-ICM, with \(\epsilon=10^{-6}\) for both. 
In the UG-ICM-based experiment, the decoder preference parameter is fixed to the machine-oriented setting ($\beta=1$) for both the teacher and student models.
Recognition performance is evaluated using YOLOv5\cite{yolov5} and Mask R-CNN\cite{mask_rcnn} for object detection, and  Mask R-CNN for instance segmentation.

We compare the proposed method with the corresponding fixed-rate ICM
model, AQVR-ICM~\cite{AQVR-ICM}, and non-adaptive quantization, which uses uniform step size scaling without adaptive control. 
For the SA-ICM-based experiment, we additionally include progressive decoding~\cite{ch-arm}
and TCM~\cite{TCM}, the base codec of SA-ICM. For the UG-ICM-based experiment, we also compare the human-oriented output of UG-ICM.

Fig.~\ref{fig:outputs} shows examples of decoded images and YOLOv5 detection results. Compared with the fixed-rate model outputs, the proposed method preserves task-relevant information at a much lower bitrate.
Figs.~\ref{fig:SA-ICM} and~\ref{fig:UG-ICM} illustrate the Rate-mAP curves
when SA-ICM and UG-ICM are used as the base models, respectively.
The proposed method provides continuous bitrate control with a single model and maintains competitive recognition performance across tasks and base architectures. These results demonstrate that importance-adaptive distillation effectively preserves task-relevant
latent components while enabling flexible rate control in a loss-agnostic manner.

\section{Conclusion}
In this paper, we proposed an importance-adaptive distillation framework
for converting fixed-rate ICM models into variable-rate models.
By using entropy-model scale parameters as an importance
cue, the proposed method adapts quantization and distillation without
requiring the original loss or downstream labels.
Experiments indicated that the proposed framework is applicable to existing ICM models with different objectives. 
The results also showed that a single adapted model maintains recognition performance
close to that of fixed-rate ICM models across a wide bitrate range.
Future work includes automatic bitrate allocation.

\section*{Acknowledgment}
This text is a part of the outcome of research performed under a Waseda University Grant for Special Research Projects (Project number: 2026R-021).

\end{document}